\documentclass[aps,prd,preprint,groupedaddress,showpacs]{revtex4}

\usepackage{epsfig}
\usepackage{graphics}

\begin{document}

\leftmargin -2cm
\def\choosen{\atopwithdelims..}
~~\\
 DESY~09--212 \hfill ISSN 0418-9833
\\December 2009

\vspace{2cm}

 \boldmath
\title{Diphoton production at Tevatron\\ in the
quasi-multi-Regge-kinematics approach} \unboldmath

  \author{\firstname{V.A.} \surname{Saleev}}
\email{saleev@ssu.samara.ru, saleev@mail.desy.de}

 \affiliation{{II.} Institut f\"ur Theoretische Physik, Universit\" at Hamburg,
Luruper Chaussee 149, 22761 Hamburg, Germany}

\affiliation{ Samara State University, Academic Pavlov Street 1,
443011 Samara, Russia}

\begin{abstract}
We study the production of prompt diphotons in the central region of
rapidity within the framework of the quasi-multi-Regge-kinematics
approach applying the hypothesis of quark and gluon Reggeization. We
describe accurately and without free parameters the experimental
data which were obtained by the CDF Collaboration at the Tevatron
Collider. It is shown that the main contribution to studied process
is given by the direct fusion of two Reggeized gluons into a photon
pair, which is described by the effective Reggeon-Reggeon to
particle-particle vertex. The contribution from the annihilation of
Reggeized quark-antiquark pair into a diphoton is also considered.
At the stage of numerical calculations we use the
Kimber-Martin-Ryskin prescription for unintegrated quark and gluon
distribution functions, with the Martin-Roberts-Stirling-Thorne
collinear parton densities for a proton as input.
\end{abstract}


\maketitle \maketitle

\section{Introduction}
\label{sec:one} The study of production of photons with large
transverse momenta in the hard interaction between two partons in
the high-energy hadron collisions, so-called prompt photon
production, provides precision tests of perturbative quantum
chromodynamics (QCD) as well as information on the parton densities
within a hadron. Also, these studies are our potential for the
observation of a new dynamical regime, namely the high-energy Regge
limit, which is characterized by the following condition
$\sqrt{S}>>\mu>>\Lambda_{QCD}$, where $\sqrt{S}$ is the total
collision energy in the center of mass reference frame,
$\Lambda_{QCD}$ is the asymptotic scale parameter of QCD, $\mu$ is
the typical energy scale of the hard interaction. At this
high-energy limit, the contribution from the partonic subprocesses
involving $t-$channel parton (quark or gluon) exchanges to the
production cross section can become dominant. In the region under
consideration, the transverse momenta of the incoming partons and
their off-shell properties can no longer be neglected, and we deal
with "Reggeized" $t-$channel partons. In this kinematics, so-called
quasi-multi-Regge kinematics, the particles (multi-Regge) or the
groups of particles (quasi-multi-Regge) produced in the collision
are strongly separated in rapidity. In our discussion, the both
photons from the photon pair are produced in the central region of
rapidity.

The quasi-multi-Regge-kinematics (QMRK) approach
\cite{FadinLipatov96} is particularly appropriate for this kind of
high-energy phenomenology. It is based on an effective quantum field
theory implemented with the non-Abelian gauge-invariant action
including fields of Reggeized gluons \cite{Lipatov95} and Reggeized
quarks \cite{LipatoVyazovsky}. In the QMRK approach we can do
calculations with Reggeized quarks
\cite{KniehlSaleevShipilova,Saleev2008} that presents an open
question in the other uncollinear factorization scheme, namely
$k_T-$factorization approach \cite{KTALL}, following which we can
operate correctly with off-shell gluons only. Recently it was shown
also, that the calculation in the next to leading order (NLO) in
strong coupling constant within the framework of QMRK approach can
be done \cite{NLO}.

Our previous studies of charmonium and bottomonium production
\cite{KniehlSaleevVasin}, open charm production
\cite{KniehlSaleevShipilova} and inclusive prompt photon production
at the DESY HERA and at the Fermilab Tevatron \cite{Saleev2008}
demonstrated the advantages of the high-energy factorization scheme
based on the QMRK approach over the collinear parton model as far as
the description of experimental data is concerned.

This paper continues our study of the inclusive prompt photon
production at Tevatron which was performed recently
\cite{Saleev2008} in the framework of the the QMRK approach. We
consider here the diphoton production at the Fermilab Tevatron in
the central region of rapidity applying the hypothesis of quark and
gluon Reggeization
\cite{FadinLipatov96,LipatoVyazovsky,FadinSherman}.

It is shown that the main contribution to studied process is given
by the direct fusion of two Reggeized gluons into a photon pair,
which is described by the effective Reggeon-Reggeon to
particle-particle vertex. The contribution from the annihilation of
Reggeized quark - antiquark pair into a diphoton is not small and it
is taken into account additionaly. We do not take into consideration
the contribution of the fragmentation mechanism of the prompt photon
production which is strongly suppressed by the isolation cone
condition \cite{Saleev2008}.

At the stage of numerical calculations we use the
Kimber-Martin-Ryskin prescription \cite{KMR} for unintegrated quark
and gluon distribution functions, with the
Martin-Roberts-Stirling-Thorne (MRST) collinear parton densities for
a proton as input \cite{MRST} .

 This paper is organized as follows.
In Sec.~\ref{sec:two}, the relevant Reggeon-Reggeon to
particle-particle effective vertices are presented and discussed. In
Sec.~\ref{sec:three}, we describe diphoton production spectra at the
Tevatron Collider. In Sec.~\ref{sec:foure}, we summarize our
conclusions.

\section{Basic Formalism}
\label{sec:two}

In the phenomenology of strong interactions at high energies, it is
necessary to describe the QCD evolution of the parton distribution
functions of colliding particles starting with some scale $\mu_0$,
which controls a non-perturbative regime, to the typical scale $\mu$
of the hard-scattering processes, which is typically of the order of
the transverse mass $M_T=\sqrt{M^2+|\vec {k}_T|^2}$ of the produced
particle with mass $M$ and transverse momentum $\vec{ k}_T$. In the
region of very high energies, in  so-called Regge limit, the typical
ratio $x=\mu/\sqrt{S}$ becomes very small, $x\ll1$. That leads to
large logarithmic contributions of the type $[\alpha_s\ln(1/x)]^n$
in the resummation procedure, which is described by the
Balitsky-Fadin-Kuraev-Lipatov (BFKL) evolution equation \cite{BFKL}
or other BFKL-like ones for unintegrated gluon (quark) distribution
functions $\Phi_{g,q}(x,|{\bf q}_T|^2,\mu^2)$. Correspondingly, in
the QMRK approach \cite{FadinLipatov96}, the initial-state
$t$-channel gluons and quarks are considered as Reggeons, or
Reggeized gluons $(R)$ and Reggeized quarks $(Q)$. They are off-mass
shell and carry finite transverse two-momenta ${\bf q}_T$ with
respect to the hadron beam from which they stem.

The advantages of the QMRK approach include: first, it uses
gauge-invariant amplitudes and is based on a factorization
hypothesis that is proven in the leading logarithmic approximation;
second, it carries over to non-leading orders in the strong-coupling
constant, as recently proven \cite{NLO}; third, it works both with
Reggeized gluons and with  Reggeized quarks. The Reggeization of
amplitudes provides the opportunity to efficiently take into account
large radiative corrections to processes in the Regge limit beyond
what is included in the collinear approximation, which is of great
practical importance. The particle Reggeization is known effect in
the high-energy quantum electrodynamics (QED) for electrons
\cite{GellMann}, and for gluons \cite{BFKL} and quarks
\cite{LipatoVyazovsky,FadinSherman} in QCD.

Recently, in Refs.~\cite{KTAntonov,LipatoVyazovsky}, the Feynman
rules for the effective theory based on the non-Abelian
gauge-invariant action including fields of Reggeized gluons
\cite{Lipatov95} and Reggeized quarks \cite{FadinSherman} were
derived for the induced and some important effective vertices. The
effective vertices for the  $2\to 2$ processes with  Reggeized
gluons in the initial state only, $C^{RR\to gg}$ and $C^{RR\to q\bar
q}$, were obtained in Refs. \cite{LipatovFadin89,BFKL}, the
effective vertices $C^{Q\bar Q\to gg}$ and $C^{RQ\to gq}$ with
Reggeized quark and  Reggeized gluon in the initial state were
obtained in Ref. \cite{LipatoVyazovsky}. For our purposes we need to
construct the effective vertices with two photons in the final
state, i.e. $C^{Q\bar Q\to \gamma \gamma}$ and $C^{RR\to \gamma
\gamma}$. The first one can be easy obtained from the above
mentioned vertices after cutting the contribution from the
three-gluon vertex and replacement the quark-gluon vertex by the
same quark-photon vertex. The effective vertex $C^{R R\to \gamma
\gamma}$ can be obtained using the Feynman rules \cite{KTAntonov}
for the Reggeized gluons and the well-known fourth-rank vacuum
polarization tensor, which corresponds to the set of quark-box
diagrams \cite{DeTollis}.

We perform calculations in the laboratory frame of the Tevatron
Collider, where proton and antiproton beams have equal energies,
$E_p=E_{\bar p}=900(980)$ GeV. We introduce two light-cone vectors
corresponding to the massless particle momenta as follows:
$P_1=E_1(1,0,0,1)$ and $P_2=E_2(1,0,0-1)$, where $E_1=E_p$ and
$E_2=E_{\bar p}$. Also we define two additional four-vectors
$(n^+)^{\mu}={P_2^{\mu}}/{E_2}$ and $(n^-)^{\mu}={P_1^{\mu}}/{E_1}$.
For any arbitrary four-momentum $k^\mu$, we define $k^{\pm}=k\cdot
n^\pm=k^\mu n_\mu^\pm$. We use the vertex functions defined in
Ref.\cite{LipatoVyazovsky}, which describe transition of the
Reggeized quark with the four-momentum $q$ and  gluon or photon with
the four-momentum $k$ into the on-shell massless quark with
four-momentum $k+q$:
\begin{eqnarray}
\gamma_\mu^{(+)}(k,q)=\gamma_\mu+\hat q
\frac{n^+_\mu}{k^+}=\gamma_\mu+ \hat q\frac{P_{2\mu}}{P_2\cdot k},\\
\gamma_\mu^{(-)}(k,q)=\gamma_\mu+\hat q
\frac{n^-_\mu}{k^-}=\gamma_\mu+ \hat q\frac{P_{1\mu}}{P_1\cdot k}.
\end{eqnarray}

The effective Reggeon-Reggeon to particle-particle vertex $C^{\bar Q
Q\to \gamma \gamma}_{\mu\nu}(q_1,q_2,k_1,k_2)$, which describes the
Reggeized quark -- Reggeized antiquark annihilation into a photon
pair in the process
\begin{eqnarray}
Q(q_1)+\bar Q(q_2)\to \gamma(k_1)+\gamma(k_2)\label{proc:QQ}
,\end{eqnarray} can be presented as follows:
\begin{eqnarray}
C^{\bar Q Q\to \gamma \gamma}_{\mu\nu}(q_1,q_2,k_1,k_2)&=&-e_q^2 e^2
\Bigl[\gamma_\nu^{(+)}(k_2,q_2)\frac{\hat q_1-\hat
k_1}{(q_2+k_2)^2}\gamma_\mu^{(-)}(-k_1,q_1)+\nonumber\\
&+&\gamma_\mu^{(+)}(k_1,q_2)\frac{\hat q_1-\hat
k_2}{(q_2+k_1)^2}\gamma_\nu^{(-)}(-k_2,q_1)+\Delta_{\mu\nu}(q_1,-q_2)\Bigr]\label{ver:QbarQ},
\end{eqnarray}
where momenta of the initial Reggeized parton are
$q_{(1,2)}=x_{(1,2)}P_{(1,2)}+q_{(1,2)T}$, and the induced term
$\Delta_{\mu\nu}(q_1,q_2)$ has the form
\begin{equation}\Delta_{\mu\nu}(q_1,q_2)=\hat q_1\frac{n^-_\mu
n^-_\nu}{k_1^-k_2^-}+\hat q_2 \frac{n^+_\mu
n^+_\nu}{k_1^+k_2^+}.\end{equation}
 The vertex (\ref{ver:QbarQ}) satisfies the
gauge-invariant condition, which reads as follows
\begin{equation}C^{\bar Q Q\to \gamma \gamma}_{\mu\nu}(q_1,q_2,k_1,k_2) k_2^\mu=
C^{\bar Q Q\to \gamma \gamma}_{\mu\nu}(q_1,q_2,k_1,k_2) k_1^\nu=0.\end{equation}

The effective gauge-invariant vertex $C_{\mu\nu}^{R R\to \gamma
\gamma}(q_1,q_2,k_1,k_2)$, which describes the direct Reggeized
gluon fusion into a diphoton in the process
\begin{eqnarray}R(q_1)+R(q_2)\to \gamma(k_1)+\gamma(k_2)\label{proc:RR},\end{eqnarray}
 can be presented as follows
\begin{eqnarray}
C_{\mu\nu}^{R R\to \gamma\gamma}(q_1,q_2,k_1,k_2)=\Pi^{(-)\alpha}_T
(q_1)\Pi^{(+)\beta}_T (q_2)G_{\alpha\beta\mu\nu}(q_1,q_2,k_1,k_2)
\label{ver:RR},
\end{eqnarray}
$G_{\alpha\beta\mu\nu}(q_1,q_2,k_1,k_2)$ is the vacuum polarization
tensor of forth rank corresponding the set of quark-box diagrams,
which are taken with the relevant color factor.

It is suitable to define the new csalar vertex $C^{R R\to
\gamma\gamma}(q_1,q_2,k_1,k_2)$, in which the summation over final
photon polarizations has been performed:
\begin{eqnarray}
C^{R R\to \gamma\gamma}(q_1,q_2,k_1,k_2)=\sum_{\lambda_1,\lambda_2}
\varepsilon^\mu(k_1,\lambda_1)\varepsilon^{\nu}(k_2,\lambda_2)C_{\mu\nu}^{R
R\to \gamma\gamma}(q_1,q_2,k_1,k_2),\label{ver:RR2}
\end{eqnarray}
where $\varepsilon^\mu(k_{1},\lambda_{1})$ and
$\varepsilon^\nu(k_{2},\lambda_{2})$ are the polarization
four-vectors of final photons. The vertex (\ref{ver:RR2}) is
presented as follows
\begin{eqnarray}
C^{R R\to \gamma\gamma}(q_1,q_2,k_1,k_2)=\Pi^{(-)\alpha}_T
(q_1)\Pi^{(+)\beta}_T (q_2)\tilde G_{\alpha\beta}(q_1,q_2,k_1,k_2)
\label{ver:RR},
\end{eqnarray}
where
$$\tilde G_{\alpha\beta}(q_1,q_2,k_1,k_2)=\sum_{\lambda_1,\lambda_2}
\varepsilon^\mu(k_1,\lambda_1)\varepsilon^{\nu}(k_2,\lambda_2)
G_{\alpha\beta\mu\nu}(q_1,q_2,k_1,k_2).$$ The exact expression for
the tensor $\tilde G_{\alpha\beta}(q_1,q_2,k_1,k_2)$ have been
obtained in the Ref.~\cite{DeTollis}. We use massless four-quark
scheme to calculate this tensor.

To obtain the vertex (\ref{ver:RR}) we operate with  projectors on
the Reggeized gluon states which can be presented in the two
equivalent forms $\Pi^{(+)\nu}_T
(q_2)=\displaystyle{\frac{q_{2T}^\nu}{|\vec q_{2T}|}}$ or
$\Pi^{(+)\nu}_T(q_2)=-\displaystyle{\frac{x_2E_2(n^{+})^\nu}{|\vec
q_{2T} |}}$ and $\Pi^{(-)\nu}_T
(q_1)=\displaystyle{\frac{q_{1T}^\nu}{|\vec q_{1T}|}}$ or
$\Pi^{(-)\nu}_T(q_1)=-\displaystyle{\frac{x_1E_1(n^{-})^\nu}{|\vec
q_{1T} |}}$. On the contrary to the definition used in
Ref.~\cite{LipatoVyazovsky}, we do not include on-shell quark
spinors in our equations for the effective vertex (\ref{ver:QbarQ}).
We also used different normalization for the projector
$\Pi^{(\pm)\mu}_T$, which reads accordingly
Ref.\cite{LipatoVyazovsky,KTAntonov} as
$\Pi^{(\pm)\mu}=(n^\pm)^\mu$. Our definition implies that the
squared Reggeized  amplitude in the QMRK approach is normalized to
the squared parton amplitude for on-shell quarks and gluons when
${\vec q}_{1T}={\vec q}_{2T}=0$.

At the next step we write the squared matrix elements of the above
mentioned Reggeized parton processes taking into account kinematical
conditions of the Tevaron Collider.

The squared matrix element for the direct diphoton  production via
annihilation (\ref{proc:QQ}) of Reggeized quark from a proton
($q_1=x_1P_1+q_{1T}$) and Reggeized antiquark from a antiproton
($q_2=x_2P_2+q_{2T}$)  is obtained from the effective vertex
(\ref{ver:QbarQ}) and it is presented as follows
\begin{eqnarray}
\overline{|M(Q_p \bar Q_{\bar p}\to \gamma \gamma)|^2}&=&
\frac{32}{3}\pi^2e_q^4\alpha^2\frac{x_1x_2}{a_1a_2b_1b_2S\hat t\hat
u}\Bigl(w_0+w_1S+w_2S^2+w_3S^3\Bigr),\label{amp:QQ}
\end{eqnarray}
where $a_1=2k_1\cdot P_2/S$, $a_2=2k_2\cdot P_2/S$, $b_1=2k_1\cdot
P_1/S$, $b_2=2k_2\cdot P_1/S$, $S=2P_1\cdot P_2$, $t_1=-q_{1T}^2$,
$t_2=-q_{2T}^2$, $\hat s=(q_1+q_2)^2$, $\hat t=(q_1-k_1)^2$, $\hat
u=(q_1-k_2)^2$, and we apply that $\hat s+\hat t +\hat u=-t_1-t_2$,
$x_1=a_1+a_2$, $x_2=b_1+b_2$,
\begin{eqnarray}
w_0=t_1t_2(t_1+t_2)-\hat t\hat u(\hat t+\hat u),
\end{eqnarray}
\begin{eqnarray}
-w_1&=&t_1t_2(a_1-a_2)(b_1-b_2)+t_2x_1(b_2\hat t+b_1\hat u)+\nonumber\\
&+&t_1x_2(a_1\hat t+a_2\hat u)+\hat t\hat
u(a_1b_1+2a_2b_1+2a_1b_2+a_2b_2),
\end{eqnarray}
\begin{eqnarray}
-w_2=b_1b_2x_1^2t_2+a_1a_2x_2^2t_1+a_1b_2\hat
t(x_1b_1+a_2b_2)+a_2b_1\hat u(a_1b_1+a_2x_2),
\end{eqnarray}
\begin{eqnarray}
-w_3=a_1a_2b_1b_2\Bigl(a_1b_2\Bigl(\frac{\hat t}{\hat
u}\Bigr)+a_2b_1\Bigl(\frac{\hat u}{\hat t}\Bigr)\Bigr).
\end{eqnarray}
When we consider the collinear approximation, in which one has
$\displaystyle{a_1=-\frac{\hat u}{x_2S}}$,
$\displaystyle{a_2=-\frac{\hat t}{x_2S}}$,
$\displaystyle{b_1=-\frac{\hat t}{x_1S}}$,
$\displaystyle{b_2=-\frac{\hat u}{x_1S}}$,
$\displaystyle{x_1x_2=\frac{\hat s}{S}}$, and $t_1=t_2=0$, the
well-known answer for the  squared on-shell parton amplitude is
obtained
\begin{eqnarray}
\overline{|M(\bar q q\to \gamma \gamma)|^2}&=&
\frac{32}{3}\pi^2e_q^4\alpha^2\Bigl(\frac{\hat t}{\hat u}+\frac{\hat
u}{\hat t}\Bigr).
\end{eqnarray}

The squared matrix element for the symmetric subprocess $\bar Q_p
Q_{\bar p} \to \gamma \gamma$ is treated very similarly, one can be
obtained from (\ref{amp:QQ}) by the replacements: $t_1
\leftrightarrow t_2$, $\hat t \leftrightarrow \hat u$, $a_1
\leftrightarrow b_1$, $a_2 \leftrightarrow b_2$, and $x_1
\leftrightarrow x_2$. It is easy to see that these replacements do
not change formula (\ref{amp:QQ}).

The squared matrix element $\overline{|M(R R\to \gamma \gamma)|^2}$
for the direct diphoton  production via the fusion of Reggeized
gluons (\ref{proc:RR})  is obtained from the effective vertex
(\ref{ver:RR}). The analytical answer for the squared matrix element
is known, see Ref.~\cite{BergerBraatenField}, in the following
approximation: $m_q=0$ and $t_{1,2}\ll \hat s,\hat t, \hat u$, where
$m_q$ is the mass of an quark in the relevant amplitude. It can be
presented in the following form:
\begin{eqnarray}
\overline{|M(R R\to \gamma
\gamma)|^2}&=&2\alpha^2\alpha_s^2\biggl(\sum_{i=1}^{n_f}e_q^2
\biggr)^2\bigl(f_1+f_2+f_3\bigr),\label{amp:RR}
\end{eqnarray}
where $n_f=4$ is the number of active quark flavors,
\begin{eqnarray}
    f_1&=&\frac{1}{8}\Biggl[\Biggl(\frac{\hat s^2+\hat t^2}{\hat u^2} \log^2\biggl(\frac{-\hat s}{\hat t}\biggr)+
    2\frac{\hat s-\hat t}{\hat u}\log\biggl(\frac{-\hat s}{\hat t}\biggr)\Biggr)^2
    +\nonumber\\
    &+&
    \Biggl(\frac{\hat s^2+\hat u^2}{\hat t^2} \log^2\biggl(\frac{-\hat s}{\hat u}\biggr)+
    2\frac{\hat s-\hat u}{\hat t}\log\biggl(\frac{-\hat s}{\hat u}\biggr)\Biggr)^2+\nonumber\\
    &+&\biggl(\frac{\hat u^2+\hat t^2}{\hat s^2}\biggl(\log^2\biggl(\frac{\hat t}{\hat u}\biggr)+\pi^2\biggr)
    +2\frac{\hat t-\hat u}{\hat s}\log(\frac{\hat t}{\hat
    u})\biggr)^2\Biggr]\nonumber\\
f_2&=&\frac{1}{2}\biggl[ \frac{\hat s^2+\hat t^2}{\hat
u^2}\log^2\biggl( \frac{-\hat s}{\hat t}\biggr)+\frac{\hat s^2+\hat
u^2}{\hat t^2}\log^2\biggl( \frac{-\hat s}{\hat
u}\biggr)+\nonumber\\
&+&\frac{\hat s-\hat t}{\hat u}\log\biggl( \frac{-\hat s}{\hat
t}\biggr)+\frac{\hat s-\hat u}{\hat t}\log\biggl( \frac{-\hat
s}{\hat u}\biggr)+\nonumber\\
&+&\frac{\hat t^2+\hat u^2}{\hat s^2}\biggl(\log^2\biggl(\frac{\hat
t}{\hat u} \biggr)+\pi^2\biggr)+2\frac{\hat t-\hat u}{\hat
s}\log\biggl(\frac{\hat t}{\hat u}\biggr)\biggr]\nonumber\\
f_3&=&\frac{\pi^2}{2}\biggl[ \biggl(\frac{\hat s^2+\hat t^2}{\hat
u^2}\log\biggl( \frac{-\hat s}{\hat u}\biggr)+\frac{\hat s-\hat
t}{\hat u}\biggr)^2+ \biggl(\frac{\hat s^2+\hat u^2}{\hat
t^2}\log\biggl( \frac{-\hat s}{\hat t}\biggr)+\frac{\hat s-\hat
u}{\hat t}\biggr)^2\biggr]+4.\nonumber
\end{eqnarray}
The formula (\ref{amp:RR}) formally coincides with the result, which
is obtained in the collinear parton model. Any way, we incorporate
the (\ref{amp:RR}) with the off-shell kinematics for the initial
Reggeized-gluons. Of course, the exact squared matrix elements with
the off-shell initial particles is needed. Our relevant calculations
for arbitrary values of $t_1$ and $t_2$ are in the progress.


\section{Prompt diphoton production at Tevatron}
\label{sec:three} During last decade the CDF and D0 Collaborations
at Tevatron obtained experimental data for the production of prompt
photons: inclusively \cite{CDFgamma,D0gamma},  in association with a
jet \cite{D0jet}, in association with a heavy quark
($c,b$)\cite{D0heavy}, and in pair \cite{CDF2gamma}. The inclusive
prompt photon production was studied in the QMRK approach in the
Ref.\cite{Saleev2008}. It was shown that the main mechanism of the
inclusive prompt photon production in  $p\bar p$ collisions is the
fusion of  Reggeized quark and  Reggeized antiquark into a photon,
via the effective Regeon-Reggeon to particle (to photon) vertex
$C^{Q\bar Q\to \gamma}$. We have described well the inclusive photon
transverse momentum spectra measured by the CDF and D0
Collaborations \cite{CDFgamma,D0gamma} in the wide region of the
photon transverse momentum and pseudorapidity.

Here we consider the production of prompt photon pair or diphoton in
the framework of the QMRK approach and compare our predictions with
the CDF data \cite{CDF2gamma}. The cross sections are measured
differentially as a function of the diphoton transverse momentum
($p_T$), the diphoton invariant mass ($M$), and the azimuthal angle
between the two photons ($\triangle\varphi$). In this experimental
analysis, the isolation condition required that the transverse
energy sum in a cone of radius $R=0.4$ (in $\varphi - \eta$ space)
about the photon direction, minus the photon energy, be less than 1
GeV. It is important to note, that the both photons in the CDF
experiment \cite{CDF2gamma} are produced in the central region of
rapidity (pseudorapidity) $|\eta_{1,2}|<0.9$. We can consider such
photons as a quasi-multi-Regge group of particles.

There are two mechanisms of prompt photon production: the production
of direct photons and the production of photons in fragmentation of
produced quarks and gluons into photons. We have obtained
\cite{Saleev2008} that the contribution of  fragmentation mechanism
is strongly suppressed by the isolation cone condition, which is
applied to the experimental data for the inclusive photon production
\cite{CDFgamma,D0gamma}. We estimate one as small, about 5 \%, and
do not take into account this contribution in the presented analysis
too.

The leading contributions in diphoton production are originating
from the  $2\to 2$ processes (\ref{proc:QQ}) and (\ref{proc:RR}).
The first one is order of $\alpha^2$, the second one is order of
$\alpha^2\alpha_s^2$.

The factorization formulas for the hadron cross sections in the QMRK
approach are the following:
\begin{eqnarray}
d\sigma(p\bar p\to \gamma\gamma X)&=&\int\frac{dx_1}{x_1}\int
\frac{d^2q_{1T}}{\pi}\int\frac{dx_2}{x_2}\int
\frac{d^2q_{2T}}{\pi}d\hat\sigma(Q\bar
Q\to\gamma\gamma)\times\nonumber\\
&&\times\left[\Phi_{Q/p}(x_1,t_1,\mu^2)\Phi_{\bar Q/\bar
p}(x_2,t_2,\mu^2)+\Phi_{\bar Q/p}(x_1,t_1,\mu^2)\Phi_{
Q/\bar p}(x_2,t_2,\mu^2)\right]\label{equ:QQ}\\
d\sigma(p\bar p\to \gamma\gamma X)&=&\int\frac{dx_1}{x_1}\int
\frac{d^2q_{1T}}{\pi}\int\frac{dx_2}{x_2}\int
\frac{d^2q_{2T}}{\pi}d\hat\sigma(RR\to\gamma\gamma)\times\nonumber\\
&&\times\Phi_{R/p}(x_1,t_1,\mu^2)\Phi_{R/\bar
p}(x_2,t_2,\mu^2)\label{equ:RR}
\end{eqnarray}
where $\Phi_{Q,R/p}(x,t,\mu^2)$ are the unintegrated  parton
distribution functions, $d\hat\sigma(RR,Q\bar Q\to\gamma\gamma)$ are
the Reggeized parton cross sections. The unintegrated  distribution
functions and the corresponding collinear $F_{q,g/p}(x,\mu^2)$
distribution functions are connected by the normalization condition
\begin{equation}
xF_{q,g/p}(x,\mu^2)=\int_0^{\mu^2} \Phi_{Q,R/p}(x,t,\mu^2)dt,
\end{equation}
that ensures the correct transition to the collinear-parton limit of
equations (\ref{equ:QQ}) and (\ref{equ:RR}). The Reggeized parton
cross sections are defined in the following manner
\begin{eqnarray}
d\hat\sigma(RR,Q\bar Q\to \gamma
\gamma)=(2\pi)^4\delta^{(4)}(q_1+q_2-k_1-k_2)\frac{\overline{|M(RR,Q\bar
Q\to \gamma \gamma)|^2}}{I_{RR,Q\bar Q}}
 \prod_{i=1}^2\frac{d^3k_i}{(2\pi)^32k_{i}^0},\label{f:sigma}
\end{eqnarray}
where $I_{RR,Q\bar Q}=2x_1x_2S$ is the flux factor of the incoming
particles, $k_1^\mu=(k_1^0,{\vec k}_{1T},k_1^z)$ and
$k_2^\mu=(k_2^0,{\vec k}_{2T},k_2^z)$ are the four-momenta of
produced photons.

Using the formulas (\ref{equ:QQ}), (\ref{equ:RR}) and
(\ref{f:sigma}), we obtain the differential cross sections for the
diphoton production, which can be written in case of the process
(\ref{proc:RR}) as follows:
\begin{eqnarray}
\frac{d\sigma(p\bar p\to \gamma\gamma
X)}{d(\triangle\varphi)}&=&\frac{1}{16\pi^3}\int dt_1\int
d\varphi_1\int dk_{1T}\int d\eta_1\int dk_{2T}\int
d\eta_2\frac{k_{1T}k_{2T}\overline{|M(RR\to\gamma\gamma)|^2}}{(x_1x_2S)^2}\nonumber\\
&& \times\Phi_{R/p}(x_1,t_1,\mu^2)\Phi_{R/\bar p}(x_2,t_2,\mu^2)\\
\frac{d\sigma(p\bar p\to \gamma\gamma
X)}{dp_T}&=&\frac{p_T}{16\pi^3}\int dt_1\int d\varphi_1\int
dk_{1T}\int d\eta_1\int dk_{2T}\int
d\eta_2\frac{\overline{|M(RR\to\gamma\gamma)|^2}}{(x_1x_2S)^2|\sin(\triangle\varphi)|}\nonumber\\
&& \times\Phi_{R/p}(x_1,t_1,\mu^2)\Phi_{R/\bar p}(x_2,t_2,\mu^2)\\
\frac{d\sigma(p\bar p\to \gamma\gamma
X)}{dM}&=&\frac{M}{16\pi^3}\int dt_1\int d\varphi_1\int dk_{1T}\int
d\eta_1\int dk_{2T}\int
d\eta_2\frac{\overline{|M(RR\to\gamma\gamma)|^2}}{(x_1x_2S)^2|\sin(\triangle\varphi)|}\nonumber\\
&& \times\Phi_{R/p}(x_1,t_1,\mu^2)\Phi_{R/\bar p}(x_2,t_2,\mu^2)
\end{eqnarray}
where $k_{1,2T}=|\vec k_{1,2T}|$, $\eta_{1,2}$ are the photon
pseudorapidities, $\varphi_1$ is the azimuthal angle between $\vec
k_{1T}$ and $\vec q_{1T}$, $x_1=(k_1^0+k_2^0+k_1^z+k_2^z)/\sqrt{S}$,
$x_2=(k_1^0+k_2^0-k_1^z-k_2^z)/\sqrt{S}$,
$k_{1,2}^0=\frac{k_{1,2T}}{2}(e^{\eta_{1,2}}+e^{-\eta_{1,2}})$,
$k_{1,2}^z=\frac{k_{1,2T}}{2}(e^{\eta_{1,2}}-e^{-\eta_{1,2}})$. The
differential cross sections for the diphoton production in case of
the  process (\ref{proc:QQ}) are written similarly. To perform the
multi-dimensional integration we use the REWIAD code from the
CERNLIB program library and we control the accuracy at the level of
2-3 percent.

\begin{figure}[ht]
\begin{center}
\includegraphics[width=.9\textwidth, clip=]{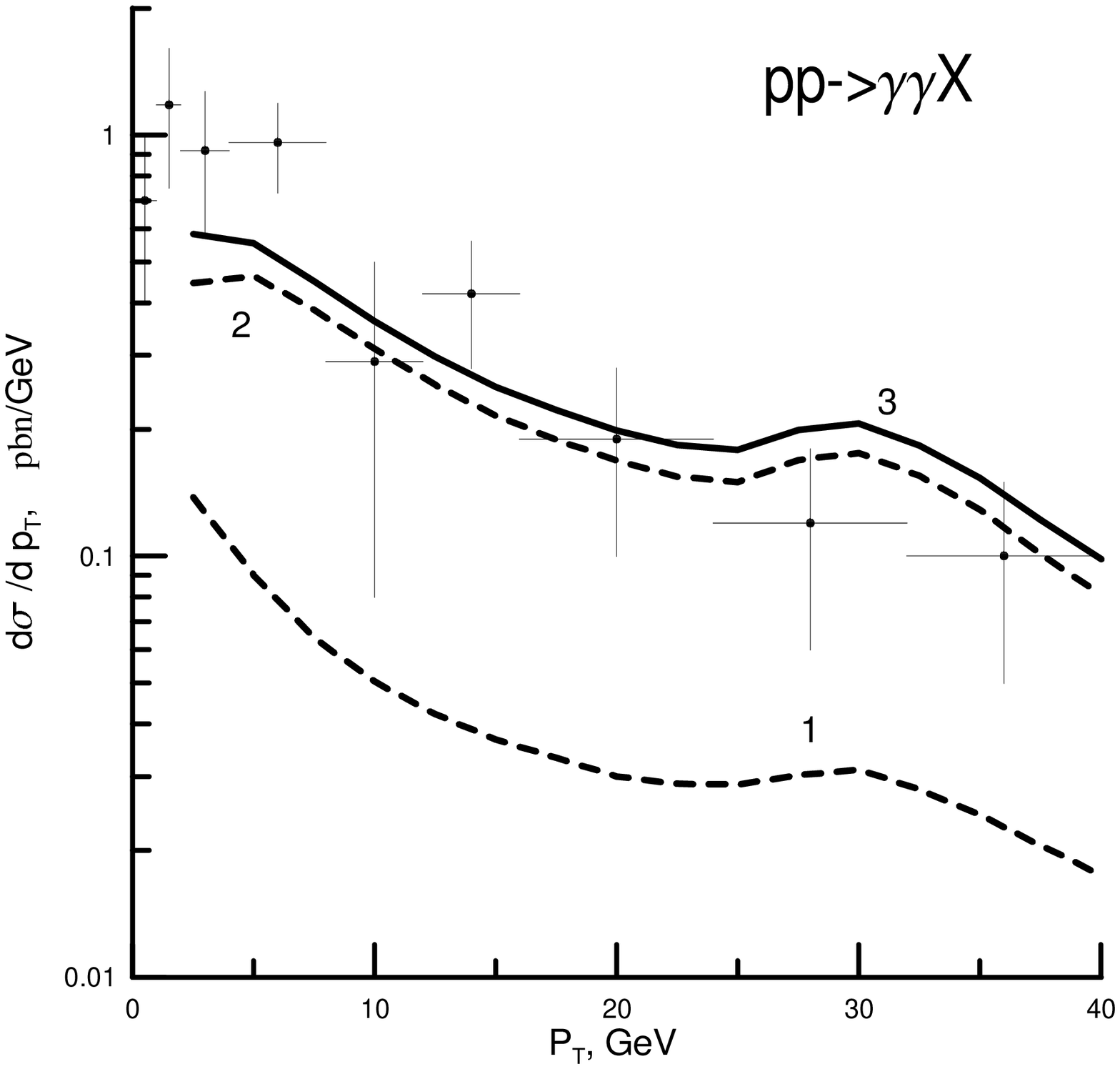}
\end{center}
\caption{The diphoton $p_T$ distribution from the CDF Collaboration
\cite{CDF2gamma}. The curve 1 is the contribution of the
 $Q\bar Q\to \gamma\gamma$ process, 2 is the contribution of the $RR\to\gamma\gamma$ process, and 3 is their sum.\label{fig:1}}
\end{figure}

\begin{figure}[ht]
\begin{center}
\includegraphics[width=.9\textwidth, clip=]{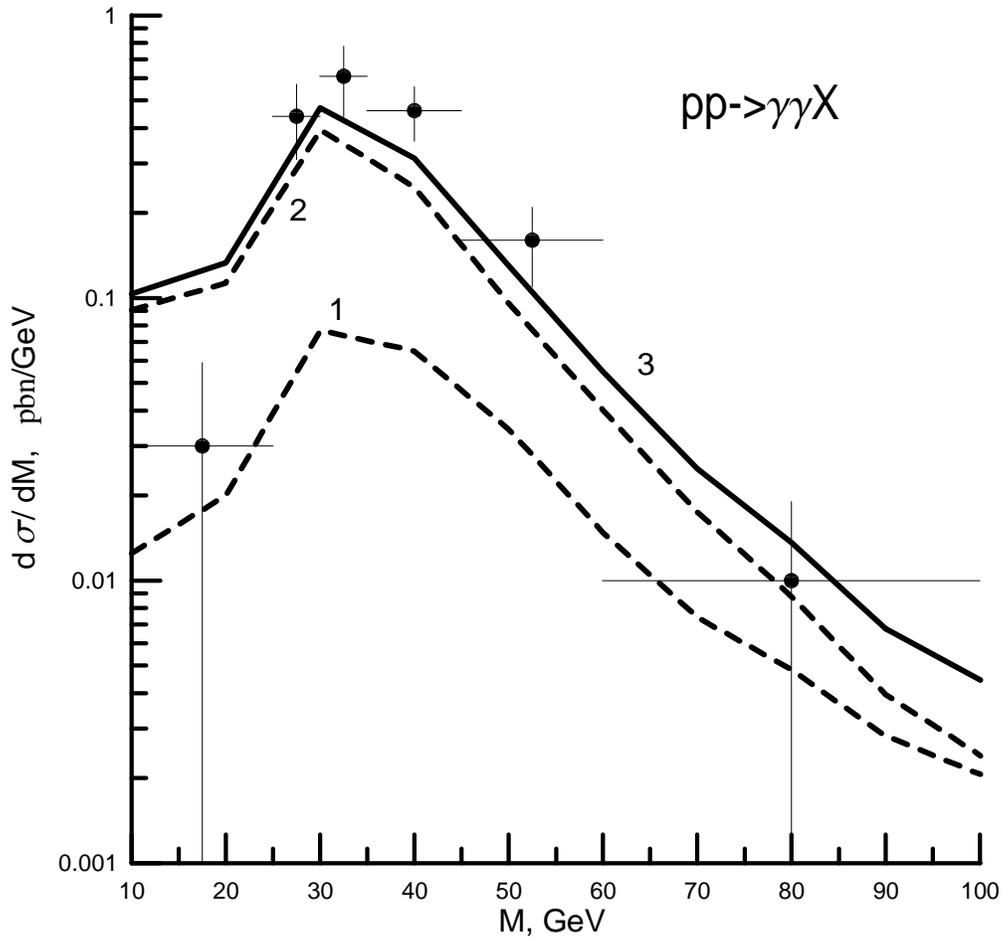}
\end{center}
\caption{The diphoton mass distribution from the CDF Collaboration
\cite{CDF2gamma}. The curves are the same as in
Fig.~1.\label{fig:2}}
\end{figure}

\begin{figure}[ht]
\begin{center}
\includegraphics[width=.9\textwidth, clip=]{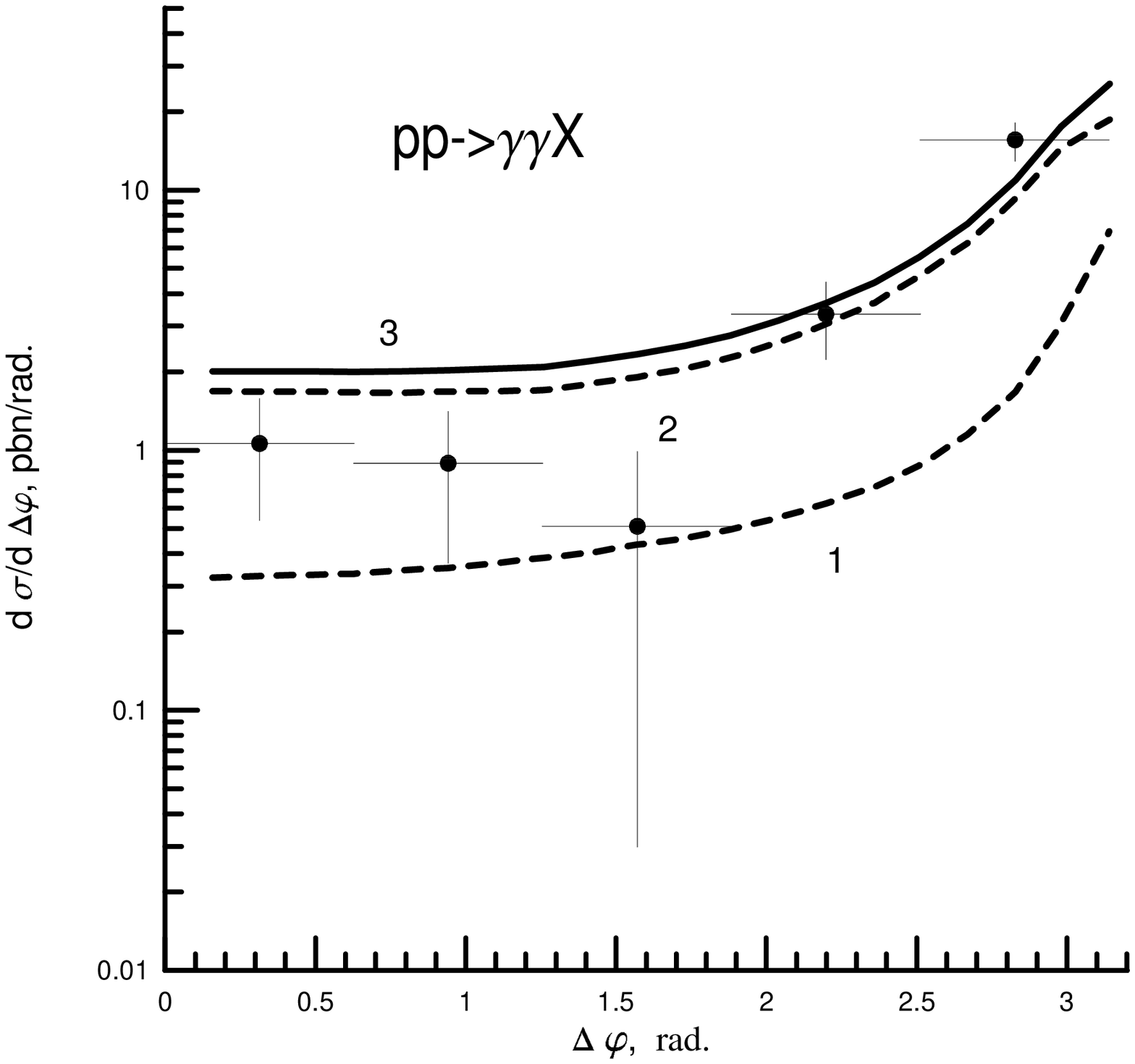}
\end{center}
\caption{The diphoton $\triangle\varphi$ distribution from the CDF
Collaboration \cite{CDF2gamma}. The curves are the same as in
Fig.~1.\label{fig:3}}
\end{figure}

We compare the results of our calculations with the experimental
data from the CDF Collaboration \cite{CDF2gamma}. The kinematic
region under consideration is defined by the following conditions
$\sqrt{S}=1960$ GeV, $|\eta_{1,2}|<0.9$, $k_{1T}>14$ GeV and
$k_{2T}>13$ GeV. The results of our calculation are shown in the
Figs. \ref{fig:1}-\ref{fig:3}. The description of the data can be
considered as good at the whole kinematic region of variables $M$,
$p_T$ and $\triangle\varphi$. It is very interesting to compare the
relative weight of different processes in connection with the
$\alpha_s$ order which they have. So, the contribution of
$RR\to\gamma\gamma$ process, which is order of $\alpha^2\alpha_s^2$,
is greater, by factor $5-10$, than the contribution of  $Q\bar
Q\to\gamma\gamma$ process, which is order of $\alpha^2$. This fact
demonstrates a dominance of the gluon induced interactions at high
energies. We see that suppression by factor $\alpha_s^2$ is smaller
that enhancement of the gluon density over the quark density in a
proton at the small $x$. The contributions of the $2\to 3$ processes
($QR\to \gamma\gamma q$ and $Q\bar Q\to\gamma\gamma g$), which are
order of $\alpha^2\alpha_s$, would be unimportant for the
theoretical approach used here. At first, the events when the
diphoton and the quark (gluon) jet are produced with close
(pseudo)rapidities are suppressed by the isolation cone condition.
At second, the events when the diphoton and the quark (gluon) jet
are produced with the large (pseudo)rapidity  gap are included in
the presented consideration via the effective vertices for Reggeized
parton interactions in the uncollinear factorization scheme, in
which we use  unintegrated over the transverse momentum parton
distribution functions.

\boldmath
\section{Conclusions}
\unboldmath \label{sec:foure} We have shown that it is possible to
describe data for the prompt diphoton production in high-energy
$p\bar p$ collisions at the Tevatron Collider in the leading order
of the QMRK approach. The scheme of our calculation bases on the
hypothesis of quark and gluon Reggeization in the hard processes at
high energy.  Our results demonstrate that the QMRK approach is a
powerful tool in the high-energy phenomenology. It is evidently that
the QMRK approach can be used also for description of the photon
plus jet production \cite{D0jet} and the photon plus heavy quark
production \cite{D0heavy} at Tevatron and the LHC Collider. Our
results for these processes will be presented in the future
publications \cite{SaleevShipilova}.

\section{Acknowledgements}
Author thanks  B.~Kniehl and L.~Lipatov for discussions of the
questions under consideration in this paper and A.~Shipilova for
help at the stage of numerical calculations. This work was supported
in part by the DAAD Grant No. A/09/03588 and by the Federal Agency
for Education of Russian Federation, Contract No. P1338.

\newpage

\end{document}